\documentclass[aps,prl,twocolumn]{revtex4}
\usepackage{amsmath, amssymb, epsfig}

\renewcommand{\vr}{{\bf{r}}}
\newcommand{\vp}{{\bf{p}}}

\newcommand{\cA}{\mathcal{A}}
\newcommand{\cB}{\mathcal{B}}

\begin{document}

\title{Loschmidt echo decay from local boundary perturbations}

\author{Arseni Goussev and Klaus Richter} \affiliation{Institut f\"ur
  Theoretische Physik, Universit\"at Regensburg, 93040 Regensburg,
  Germany}

\date{\today}

\begin{abstract}
  We investigate the sensitivity of the time evolution of
  semiclassical wave packets in two-dimensional chaotic billiards with
  respect to local perturbations of their boundaries. For this
  purpose, we address, analytically and numerically, the time decay of
  the Loschmidt echo (LE). We find the LE to decay exponentially in
  time, with the rate equal to the classical escape rate from an open
  billiard obtained from the original one by removing the
  perturbation-affected region of its boundary. Finally, we propose a
  principal scheme for the experimental observation of the LE decay.
\end{abstract}

\maketitle

The study of the sensitivity of the quantum dynamics to perturbations
of the system's Hamiltonian is one of the important objectives of the
field of Quantum Chaos. An essential concept here is the {\it
  Loschmidt echo} (LE), also known as {\it fidelity}, that was first
introduced by Peres \cite{peres} and has been widely discussed in the
literature since then \cite{review}. The LE, $M(t)$, is defined as an
overlap of the quantum state $e^{-iHt/\hbar} |\phi_0\rangle$ obtained from an
initial state $|\phi_0\rangle$ in the course of its evolution through a time
$t$ under a Hamiltonian $H$, with the state $e^{-i\tilde{H}t/\hbar} |\phi_0\rangle$
that results from the same initial state by evolving the latter
through the same time, but under a perturbed Hamiltonian $\tilde{H}$
different from $H$:
\begin{equation}
  M(t) = \left| \langle\phi_0| e^{i\tilde{H}t/\hbar} e^{-iHt/\hbar} |\phi_0\rangle \right|^2.
\label{1}
\end{equation}
It can be also interpreted as the overlap of the initial state $|\phi_0\rangle$
and the state obtained by first propagating $|\phi_0\rangle$ through the time
$t$ under the Hamiltonian $H$, and then through the time $-t$ under
$\tilde{H}$. The LE equals unity at $t=0$, and typically decays
further in time.

Jalabert and Pastawski have analytically discovered \cite{jalab} that
in a quantum system, with a chaotic classical counterpart, Hamiltonian
perturbations (sufficiently week not to affect the geometry of
classical trajectories, but strong enough to significantly modify
their actions) result in the exponential decay of the {\it average} LE
$\overline{M(t)}$, where the averaging is performed over an ensemble
of initial states or system Hamiltonians: $\overline{M(t)} \sim e^{-\lambda
  t}$. The decay rate $\lambda$ equals the average Lyapunov exponent of the
classical system.  This decay regime, known as the Lyapunov regime,
provides a strong, appealing connection between classical and quantum
chaos, and is supported by extensive numerical simulations
\cite{cucch-3}. For discussion of other decay regimes consult
Ref.~\cite{review}.

\begin{figure}[h]
\centerline{\epsfig{figure=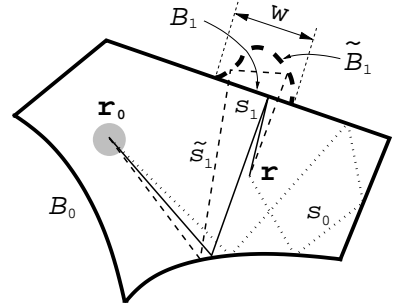,width=2.0in}}
\caption{An unperturbed, chaotic billiard (solid-line), together with
  the perturbation (dashed-line). The boundary of the unperturbed
  billiard consists of two segments, $\mathcal{B}_0$ and
  $\mathcal{B}_1$. The perturbation replaces the latter segment by
  $\tilde{\mathcal{B}}_1$, rendering the perturbed billiard to be
  bounded by $\mathcal{B}_0$ and $\tilde{\mathcal{B}}_1$. The initial
  Gaussian wave packet is centered at $\vr_0$. Three possible types of
  trajectories, $s_0$, $s_1$ and $\tilde{s}_1$, leading from $\vr_0$
  to another point $\vr$, are shown.}
\label{fig-1}
\end{figure}

\begin{figure}[h]
\centerline{\epsfig{figure=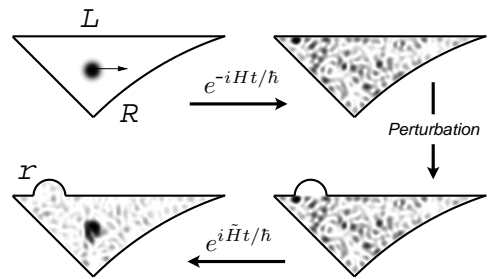,width=2.5in}}
\caption{Forward-time wave packet evolution in the unperturbed DD
  billiard, followed by the reversed-time evolution in the perturbed
  billiard. The initial Gaussian wave packet is characterized by the
  size $\sigma = 12$ and de Broglie wavelength $\lambdabar = 15/\pi$; the
  arrow shows the momentum direction of the initial wave packet. The
  DD billiard is characterized by $L = 400$ and $R = 200 \sqrt{10}$.
  The perturbation is defined by $w = 60$ and $r = 30$. The
  propagation time corresponds to approximately 10 collisions of the
  classical particle.}
\label{fig-2}
\end{figure}

In this paper we report a new regime for the time decay of the {\it
  unaveraged}, individual LE for a semiclassical wave packet evolving
  in a two-dimensional billiard that is chaotic in the classical
  limit.  We consider the general class of {\it strong} perturbations
  of the Hamiltonian that locally modify the billiard's boundary: the
  perturbation only affects a boundary segment of length $w$ small
  compared to the perimeter $P$, see Figs.~\ref{fig-1} and
  \ref{fig-2}. Both $w$ and the perturbation length scale in the
  direction perpendicular to the boundary are considered to be much
  larger than the de Broglie wavelength $\lambdabar$, so that the
  perturbation significantly modifies trajectories of the underlying
  classical system, see Fig.~\ref{fig-1}. Our analytical calculations,
  confirmed by results of numerical simulations, show that the LE in
  such a system follows the exponential decay $M(t) \sim e^{-2\gamma
  t}$, with $\gamma$ being the rate at which classical particles would
  escape from an {\it open} billiard obtained from the original,
  unperturbed billiard by removing the perturbation-affected boundary
  segment. The LE decay is independent of the shape of a particular
  boundary perturbation, and only depends on the length of the
  perturbation region. Furthermore, our numerical analysis shows that
  for certain choices of system parameters the exponential decay
  persists for times $t$ even longer than the Heisenberg time
  $t_\mathrm{H}$.

We proceed by considering a Gaussian wave packet,
\begin{equation}
  \phi_0(\vr) = \frac{1}{\sqrt{\pi} \sigma} \exp
  \left[ \frac{i}{\hbar} \: \vp_0 \cdot (\vr-\vr_0) -
    \frac{(\vr-\vr_0)^2}{2\sigma^2} \right],
\label{2}
\end{equation}
centered at a point $\vr_0$ inside the domain $\cA$ of a
two-dimensional chaotic billiard (e.g. the solid-line boundary in
Fig.~\ref{fig-1}), and characterized by an average momentum $\vp_0$
that defines the de Broglie wavelength of the moving particle,
$\lambdabar = \hbar/|\vp_0|$. The dispersion $\sigma$ is assumed to be
sufficiently small for the normalization integral $\int_\cA d\vr
|\phi_0(\vr)|^2$ to be close to unity. We let the wave packet evolve
inside the billiard through a time $t$ according to the time-dependent
Schr\"odinger equation with hard-wall (Dirichlet) boundary conditions.
This evolution yields the wave function $\phi_t(\vr) =
\langle\vr|e^{-iHt/\hbar}|\phi_0\rangle$, where $H$ stands for the Hamiltonian of the
billiard. Then, we consider a perturbed billiard obtained from the
original one by modifying the shape of a small segment of its
boundary. Fig.~\ref{fig-1} illustrates the perturbation: the
unperturbed billiard is bounded by segments $\cB_0$ and $\cB_1$,
whereas the boundary of the perturbed billiard is composed of $\cB_0$
and $\tilde{\cB}_1$. The perturbation, $\cB_1 \to \tilde{\cB}_1$, is
assumed to be such that the domain $\tilde{\cA}$ of the perturbed
billiard entirely contains the domain $\cA$ of the unperturbed one.
The time evolution of the initial wave packet, Eq.~(\ref{2}), inside
the perturbed billiard results to $\tilde{\phi}_t(\vr) =
\langle\vr|e^{-i\tilde{H}t/\hbar}|\phi_0\rangle$, with $\tilde{H}$ being the Hamiltonian
of the perturbed billiard. Then, the LE, defined in Eq.~(\ref{1}),
reads
\begin{equation}
M(t) = \left| \int_\cA d\vr \tilde{\phi}_t^*(\vr) \phi_t(\vr) \right|^2,
\label{3}
\end{equation}
where the asterisk denotes complex conjugation.

We now present a semiclassical calculation of the overlap integral
Eq.~(\ref{3}). As the starting point we take the expression
\cite{jalab,cucch-3} for the time evolution of the small (such that
$\sigma$ is much smaller than the characteristic length scale of the
billiard) Gaussian wave packet, defined by Eq.~(\ref{2}):
\begin{equation}
  \phi_t(\vr) \approx (4\pi\sigma^2)^{1/2} \sum_{s(\vr,\vr_0, \, t)} \! \mathcal{K}_s(\vr,\vr_0,t)
  \: e^{-\sigma^2 (\vp_s-\vp_0)^2 / {2\hbar^2}}.
\label{4}
\end{equation}
This expression is obtained by applying the semiclassical Van Vleck
propagator \cite{brack}, with the action linearized in the
vicinity of the wave packet center $\vr_0$, to the wave packet
$\phi_0(\vr)$. Here, the sum goes over all possible trajectories
$s(\vr,\vr_0,t)$ of a classical particle inside the unperturbed
billiard leading from the point $\vr_0$ to the point $\vr$ in time $t$
(e.g. trajectories $s_0$ and $s_1$ in Fig.~\ref{fig-1}), and
\begin{equation}
  \mathcal{K}_s(\vr,\vr_0, t) = \frac{\sqrt{D_s}}{2\pi i \hbar} 
  \exp\left( \frac{i}{\hbar} S_s(\vr,\vr_0,t) - i \frac{\pi \nu_s}{2}
  \right),
\label{5}
\end{equation}
where $S_s(\vr,\vr_0,t)$ denotes the classical action along the path
$s$. In a hard-wall billiard $S_s(\vr,\vr_0,t) = (m/2t)
L_s^2(\vr,\vr_0)$, where $L_s(\vr,\vr_0)$ is the length of the
trajectory $s$, and $m$ is the mass of the moving particle. In
Eq.~(\ref{5}), $D_s = |\det(-\partial^2S_s / \partial\vr \partial\vr_0)|$ is the Van Vleck
determinant, and $\nu_s$ is an index equal to twice the number of
collisions with the hard-wall billiard boundary that a particle,
traveling along $s$, experiences during time $t$ \cite{gasp}. In
Eq.~(\ref{4}), $\vp_s = -\partial S_s(\vr,\vr_0,t) / \partial\vr_0$ stands for the
initial momentum of a particle on the trajectory $s$. The expression
for the time-dependent wave function $\tilde{\phi}_t(\vr)$ is obtained
from Eq.~(\ref{4}) by replacing the trajectories $s(\vr,\vr_0,t)$ by
paths $\tilde{s}(\vr,\vr_0,t)$, that lead from $\vr_0$ to $\vr$ in
time $t$ within the boundaries of the perturbed billiard (e.g.
trajectories $s_0$ and $\tilde{s}_1$ in Fig.~\ref{fig-1}).

The wave functions of the unperturbed and perturbed billiards at a
point $\vr \in \cA$ can be written as
\begin{equation}
\begin{split}
  \phi_t(\vr) &= \phi_t^{(0)}(\vr) + \phi_t^{(1)}(\vr),\\
  \tilde{\phi}_t(\vr) &= \phi_t^{(0)}(\vr) + \tilde{\phi}_t^{(1)}(\vr),
\end{split}
\label{6}
\end{equation}
where $\phi_t^{(0)}(\vr)$ is given by Eq.~(\ref{4}) with the sum in the
RHS involving only trajectories $s_0$, which scatter only off the part
of the boundary, $\cB_0$, that stays unaffected by the perturbation,
see Fig.~\ref{fig-1}. On the other hand, the wave function $\phi_t^{(1)}$
( $\tilde{\phi}_t^{(1)}$ ) involves only such trajectories $s_1$ (
$\tilde{s}_1$ ) that undergo at least one collision with the
perturbation-affected region, $\cB_1$ ( $\tilde{\cB}_1$ ), see
Fig.~\ref{fig-1}. The LE integral in Eq.~(\ref{3}) has now four
contributions:
\begin{equation}
\begin{split}
  \int_\cA d\vr \tilde{\phi}_t^* \phi_t &= \int_\cA d\vr \left| \phi_t^{(0)}
  \right|^2 + \int_\cA d\vr \left[\phi_t^{(0)}\right]^* \phi_t^{(1)}\\
  & + \int_\mathcal{A} d\vr \left[\tilde{\phi}_t^{(1)}\right]^* \phi_t^{(0)} +
  \int_\cA d\vr \left[\tilde{\phi}_t^{(1)}\right]^* \phi_t^{(1)}.
\end{split}
\label{7}
\end{equation}
We argue that the dominant contribution to the LE overlap comes from
the first integral in the RHS of the last equation.  Indeed, all the
integrands in Eq.~(\ref{7}) contain the factor $\exp [ i(S_s-S_{s'})/\hbar
- i\pi(\nu_s-\nu_{s'})/2 ]$, where the trajectory $s$ is either of the type
$s_0$ or $s_1$, and $s'$ is either of the type $s_0$ or $\tilde{s}_1$,
see Fig.~\ref{fig-1}. An integral vanishes if there is no correlation
between $s$ and $s'$, since the corresponding integrand is a rapidly
oscillating function of $\vr$.  This is indeed the case for the last
two integrals: they involve such trajectory pairs $(s,s')$ that $s$ is
of the type $s_0$ or $s_1$, and $s'$ is of the type $\tilde{s}_1$, so
that the absence of correlations within such pairs is guaranteed by
the fact that the scale of the boundary deformation is much larger
than $\lambdabar$. Then, we restrict ourselves to the {\it diagonal
  approximation}, in which only the trajectory pairs with $s=s'$
survive the integration over $\vr$. The second integral in the RHS of
Eq.~(\ref{7}) only contains the trajectory pairs of the type
$(s_0,s_1)$, and, therefore, vanishes in the diagonal approximation.
Thus, the only non-vanishing contribution reads
\begin{equation}
\begin{split}
  \int_\cA d\vr \left| \phi_t^{(0)} \right|^2 &\approx \frac{\sigma^2}{\pi\hbar^2} \int_\cA d\vr
  \sum_{s_0} D_{s_0} \exp\left[-\frac{\sigma^2}{\hbar^2} \left( \vp_{s_0}-\vp_0
    \right)^2 \right]\\ &\approx \int_{\mathcal{P}_t(\cA)} d\vp \:
  \frac{\sigma^2}{\pi\hbar^2} \exp\left[-\frac{\sigma^2}{\hbar^2} \left( \vp-\vp_0
    \right)^2 \right],
\end{split}
\label{8}
\end{equation}
where $D_{s_0} = |\det(\partial\vp_{s_0}/\partial\vr)|$, with $\vp_{s_0}$ being the
initial momentum on the trajectory $s_0(\vr,\vr_0,t)$, serves as the
Jacobian of the transformation from the space of final positions $\vr
\in \cA$ to the space of initial momenta $\vp \in \mathcal{P}_t(\cA)$.
Here, $\mathcal{P}_t(\cA)$ is the set of all momenta $\vp$ such that a
trajectory, starting from the phase-space point $(\vr_0,\vp)$, arrives
at a coordinate point $\vr \in \cA$ after the time $t$, while undergoing
collisions only with the boundary $\cB_0$ (and, thus, avoiding
$\cB_1$), see Fig.~\ref{fig-1}. Thus, $\int_\cA d\vr | \phi_t^{(0)} |^2$ is
merely the probability that a classical particle, with the initial
momentum sampled from the Gaussian distribution, experiences no
collisions with $\cB_1$ during the time $t$. Therefore, if the
boundary segment $\cB_1$ is removed, this integral corresponds to the
{\it survival probability} of the classical particle in the resulting
open billiard. In chaotic billiards the survival probability decays
exponentially \cite{bauer} as $e^{-\gamma t}$, with the escape rate $\gamma$
given by
\begin{equation}
\gamma = v \frac{w}{\pi A},
\label{9}
\end{equation}
where $v=|\vp_0|/m$ is the particle's velocity, and $A$ stands for the
area of the billiard. Equation~(\ref{9}) assumes that the
characteristic escape time $1/\gamma$ is much longer that the average
free flight time $t_\mathrm{f}$. In chaotic billiards the latter is
given by \cite{niels} $t_f = \pi A / v P$, where $P$ is the perimeter
of the billiard. Condition $t_\mathrm{f} \ll 1/\gamma$ is equivalent
to $w \ll P$.

In accordance with Eqs.~(\ref{3}) and (\ref{7}) the LE decays as
\begin{equation}
M(t) \sim \exp(-2\gamma t).
\label{10}
\end{equation}
Equation~(\ref{10}) constitutes the central result of the paper.
Together with Eq.~(\ref{9}) it shows that for a given billiard the LE
merely depends on the length $w$ of the boundary segment affected by
the perturbation and on the de Broglie wavelength $\lambdabar = \hbar/mv$.
{\it It is independent of the shape and area of the boundary
  perturbation, as well as of the position, size and momentum
  direction of the initial wave packet.} (We exclude initial
conditions for which the wave packet interacts with the perturbation
before having considerably explored the allowed phase space.)

The decay rate $\gamma$, and thus the LE, are also related to classical
properties of the chaotic set of periodic trajectories unaffected by
the boundary perturbation, i.e. to properties of the chaotic repellor
of the open billiard \cite{kantz}:
\begin{equation}
\gamma = \lambda_\mathrm{r} - h_\mathrm{KS},
\label{11}
\end{equation}
where $\lambda_\mathrm{r}$ is the average Lyapunov exponent of the repellor,
and $h_\mathrm{KS}$ is its Kolmogorov-Sinai entropy. Thus,
Eqs.~(\ref{10}) and (\ref{11}) provide an interesting link between
classical and quantum chaos.

In order to verify the analytical predictions we simulated the
dynamics of a Gaussian wave packet inside a desymmetrized diamond (DD)
billiard, defined as the fundamental domain of the area confined by
four intersecting disks centered at the vertices of a
square. According to the theorem of Ref.~\cite{szasz} the DD billiard
is chaotic in the classical limit. It can be characterized by the disk
radius $R$, and the length $L$ of the longest straight segment of the
boundary, see Fig.~\ref{fig-2}. We consider the Hamiltonian
perturbation that replaces a straight segment of length $w$ of the
boundary of the unperturbed billiard by an arc of radius $r$, see
Fig.~\ref{fig-2}. In general, $w \leq 2r$.

To simulate the time evolution of the wave packet inside the billiard
we utilize the Trotter-Suzuki algorithm \cite{raedt}.
Figure~\ref{fig-2} illustrates the time evolution of a Gaussian wave
packet in the DD billiard followed by the time-reversed evolution
inside the perturbed billiard. The parameters characterizing the
system are $L = 400$, $R = 200 \sqrt{10}$, $w = 60$ and $r = 30$. The
Gaussian wave packet is parametrized by $\sigma = 12$, $\lambdabar = 15/\pi$;
the arrow shows the momentum direction of the initial wave packet. The
evolution time $t$ in Fig.~\ref{fig-2} corresponds to some 10 free
flight times of the corresponding classical particle, i.e. $t = 10
t_\mathrm{f}$.

\begin{figure}[h]
\centerline{\epsfig{figure=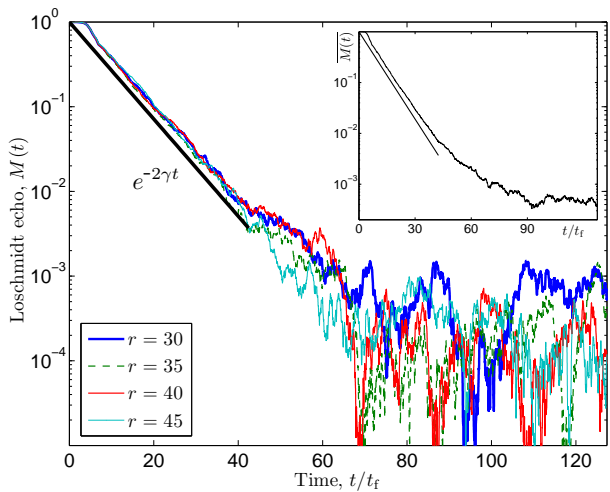,width=3.2in}}
\caption{(Color online) The Loschmidt echo (LE) decay in the DD
  billiard for four different values of the curvature radius $r$ of
  the arc perturbation. The width of the perturbation region is fixed,
  $w = 60$. The other system parameters are $L = 400$, $R = 200
  \sqrt{10}$, $\sigma = 3$, and $\lambdabar = 5/\pi$.  The solid straight
  line gives the trend of the $\exp(-2\gamma t)$ decay, with $\gamma$ given by
  Eq.~(\ref{9}).  The inset presents the decay of the average LE, with
  averaging performed over individual LE curves corresponding to
  different values of $r$.}
\label{fig-3}
\end{figure}

Figure~\ref{fig-3} shows the time dependence of the LE computed for
the DD billiard system characterized by $L = 400$, $R = 200
\sqrt{10}$, $\sigma = 3$, and $\lambdabar = 5/\pi$. The initial momentum
direction is the same as in Fig.~\ref{fig-2}. The different LE decay
curves correspond to different shapes of the local boundary
perturbation: the width of the perturbation region stays fixed, $w =
60$, and the curvature radius of the perturbation arc takes the values
$r = 30; \, 35; \, 40; \, 45$. In all four cases the LE displays the
exponential decay for times $t$ up to
40$t_\mathrm{f}$-45$t_\mathrm{f}$ followed by LE fluctuations around a
saturation value, $M_\mathrm{s}$. The thick solid straight line shows
the trend of the $e^{-2\gamma t}$ exponential decay, with $\gamma$ given by
Eq.~(\ref{9}). One can see strong agreement between the numerical and
analytical LE decay rates. We have also verified numerically that the
LE decay rate is independent of the momentum direction of the initial
wave packet.

The inset in Fig.~\ref{fig-3} presents the time decay of the average
LE $\overline{M(t)}$, with the averaging performed over 16 individual
decay curves $M(t)$ corresponding to different values of the arc
radius $r$, ranging from $r = 30$ to $45$. The saturation mechanism
for the LE decay was first proposed by Peres \cite{peres} and later
discussed in Ref.~\cite{cucch-3}. The LE saturates at a value
$M_\mathrm{s}$ inversely proportional to the number $N$ of energy
levels significantly represented in the initial state. If the areas of
the unperturbed and perturbed billiards are relatively close, then $N
\approx A/\sigma^2$ and $M_\mathrm{s} \sim \sigma^2/A$. (We have verified the latter
relation by computing the LE saturation value for billiards of
different area.) Thus, one might expect the exponential decay of the
LE to persist for times $t \lesssim t_\mathrm{s}$, with the saturation time
$t_\mathrm{s} = (1/2\gamma v) \ln N$. The latter can be longer than the
Heisenberg time $t_\mathrm{H} = A/2\pi \lambdabar v$ for a system with
sufficiently large effective Hilbert space, since
$t_\mathrm{s}/t_\mathrm{H} \sim (\lambdabar/w) \ln N$. Indeed, for the
system corresponding to Fig.~\ref{fig-3} one has $t_\mathrm{H} \approx 29
t_\mathrm{f}$, whereas the exponential decay persists for times $t <
40 t_\mathrm{f}$.

Finally, we sketch a principal experimental scheme for measuring the
LE decay regime proposed in this paper. Consider a two-dimensional,
AlGaAs-GaAs heterojunction-based ballistic cavity with the shape of a
chaotic billiard, e.g.  Fig.~\ref{fig-1}. Let the initial electron
state to be given by $| \Psi_0 \rangle = | \phi_0 \rangle \otimes | \chi \rangle$, where $| \phi_0 \rangle$ is
the spatial part defined by Eq.~(\ref{2}), and $| \chi \rangle = 2^{-1/2}
\left( | \! \uparrow \rangle + | \!  \downarrow \rangle \right)$ represents a spin-1/2 state.
Here, $| \! \uparrow \rangle$ and $| \! \downarrow \rangle$ are the eigenstates of the
spin-projection operator in the $z$-direction, perpendicular to the
billiard plane. Then, $| \chi \rangle$ is the eigenstate of the spin-projection
operator $s_x = \frac{\hbar}{2} \sigma_x$, with $\sigma_x = | \! \uparrow \rangle \langle \downarrow \! | + | \!
\downarrow \rangle \langle \uparrow \! |$, in some $x$-direction, fixed in the billiard plane.
Suppose now that a half-metallic ferromagnet, magnetized in the
$z$-direction, is attached to the boundary of the ballistic cavity.
(One may consider the region bounded by $\cB_0$ and $\cB_1$ in
Fig.~\ref{fig-1} to represent the ballistic cavity, and the region
bounded by $\cB_1$ and $\tilde{\cB}_1$ to represent the ferromagnet.)
Then the ferromagnet-cavity interface will reflect the $| \! \uparrow
\rangle$-component of the state, but will transmit the $| \! \downarrow \rangle$-component.
As a result, the two components will evolve under two different
spatial Hamiltonians, $H$ and $\tilde{H}$, corresponding to the
geometry of the ballistic cavity and the geometry of the
cavity-ferromagnet compound respectively. Then $| \Psi_0 \rangle$ will evolve
to
\begin{equation}
  | \Psi_t \rangle = \frac{1}{\sqrt{2}} \left[ e^{-iHt/\hbar} | \phi_0
    \rangle \otimes | \! \uparrow \rangle + e^{-i\tilde{H}t/\hbar} |
    \phi_0 \rangle \otimes | \! \downarrow \rangle \right].
\label{12}
\end{equation}
The expectation value of the projection of the spin in the
$x$-direction is related to the LE overlap by
\begin{equation}
\bar{s}_x(t) \equiv \langle \Psi_t | s_x | \Psi_t \rangle =
\frac{\hbar}{2} \mathrm{Re} \langle\phi_0| e^{i\tilde{H}t/\hbar}
e^{-iHt/\hbar} |\phi_0\rangle,
\label{13}
\end{equation}
where $\mathrm{Re}$ denotes the real part. As we have shown above,
this overlap is real and decays exponentially in time. Therefore, the
average spin projection in the $x$-direction will also relax
exponentially with time, i.e. $\bar{s}_x(t) \sim \frac{\hbar}{2} \exp(-\gamma t)$,
with the relaxation rate $\gamma$ determined by Eq.~(\ref{9}). This result
provides a link between the spin relaxation in chaotic, mesoscopic
structures \cite{been} and the LE decay due to local boundary
perturbations.

The authors would like to thank Inanc Adagideli, Arnd B\"acker, Fernando
Cucchietti, Philippe Jacquod, Thomas Seligman, and Oleg Zaitsev for
helpful conversations. AG acknowledges the Alexander von Humboldt
Foundation (Germany), and KR acknowledges the Deutsche
Forschungsgemeinschaft (DFG) for support of the project.

\end{document}